\documentclass[11pt,notref,notcite]{article}
\usepackage{amsfonts}
\usepackage{cite}
\usepackage{graphics,amsmath,amssymb,amsthm,accents}
\def \h#1{\widehat{#1}}
\def \t#1{\widetilde{#1}}

\def \th#1{\widehat{\widetilde{#1}}}

\numberwithin{equation}{section}
\newtheorem{prop}{Proposition}
\newcommand{\balpha}{\boldsymbol{\alpha}}

\newcommand{\bft}{\mathbf{t}}

\title{Conservation laws of some lattice equations}

\author{Jun-wei Cheng\footnote{Email: chengjw2004@yahoo.com.cn},~~
Da-jun Zhang\footnote{Corresponding author.
E-mail: djzhang@staff.shu.edu.cn} \\
{\small\it Department of Mathematics, Shanghai University, Shanghai 200444,  P.R. China}}

\begin{document}

\maketitle

\begin{abstract}
We derive infinitely many conservation laws for some
multi-dimensionally consistent lattice equations from their Lax
pairs. These lattice equations are the Nijhoff-Quispel-Capel
equation, lattice Boussinesq equation, lattice nonlinear
Schr\"{o}dinger equation, modified lattice Boussinesq equation,
Hietarinta's Boussinesq-type equations, Schwarzian lattice
Boussinesq equation and Toda-modified lattice Boussinesq equation.


\end{abstract}

\section{Introduction}

In recent years the study of integrable partial  difference
equations  has progressed rapidly. The property of multi-dimensional
consistency\cite{Nijhoff-MDC,BS-2002,ABS-2003} acts as an important
role in the research of discrete integrable systems. By this
property as a criteria and through  searching approaches many
multi-dimensionally consistent lattice equations are
found\cite{ABS-2003,Hietarinta-CAC,Hietarinta-Bous}. For such
equations one can easily write out their B\"acklund transformations
and Lax pairs, which have been used to derive solutions and
conservation laws (e.g.
\cite{Soliton-Q4,Soliton-Q3,Hietarinta-ABS,Hietarinta-Bou,Hie-B2,rs-2009,Gardner-1}).

Possessing infinitely many conservation laws is one of the important
characters of integrable systems. For discrete integrable systems,
many methods have been developed to find infinitely many
conservation laws\cite{NQC-cls-2006,rs-2009,Gardner-1,x-2011,x-2012}. Recently,
we proposed an approach to derive infinitely many conservation laws
for the Adler-Bobenko-Suris (ABS)\cite{ABS-2003} lattice equations
from their Lax pairs\cite{zcs-cls-ABS}. In this paper we will apply
the same method to some multi-component multi-dimensionally
consistent lattice equations. We will first in the next section,
taking the Nijhoff-Quispel-Capel (NQC) equation and discrete
Boussinesq (DBSQ) equation as examples, describe our approach. Then
in Sec.3 we derive conservation laws for lattice nonlinear
Schr\"{o}dinger equation, modified lattice Boussinesq equation,
Hietarinta's Boussinesq-type equations, Schwarzian lattice
Boussinesq equation and Toda-modified lattice Boussinesq equation.
We use Lax pairs collected in Ref.\cite{Bridgman-Lax}.

\section{Conservation laws of the NQC equation and DBSQ equation}\label{sec:2}

Let us take the NQC equation and DBSQ equation as examples to review the approach that we used in \cite{zcs-cls-ABS}
for deriving conservation laws. Conservation laws of these two equations have also been considered in \cite{NQC-cls-2006} and \cite{x-2012}
through direct approach and symmetry approach.

\subsection{The NQC equation}

Consider a quadrilateral lattice equation
\begin{equation}\label{quad-eq}
Q(u,\t u,\h u,\th u , p,q)=0,
\end{equation}
where
\[u=u(n,m),~~\t{u}=E_nu=u(n+1,m),~~\h{u}=E_mu=u(n,m+1),~~\th u=u(n+1,m+1),\]
$E_n$ and $E_m$ respectively serve as  shift operators in direction
$n$ and $m$, $p$ and $q$ are spacing parameters of direction $n$ and
$m,$ respectively. A conservation law of equation \eqref{quad-eq} is defined by
\begin{equation}
\Delta_mF(u)=\Delta_nJ(u),
\end{equation}
where $\Delta_m=E_m-1,~\Delta_n=E_n-1$, and $u$ is a generic solution to \eqref{quad-eq}.

The NQC equation is\cite{NQC-1983,Nijhoff-ABS}
\begin{equation}\label{NQC}
\big[(p-\alpha)u-(p+\beta)\t u\big]\big[(p-\beta)\h u-(p+\alpha)\th
u\big]-\big[(q-\alpha)u-(q+\beta)\h u\big]\big[(q-\beta)\t
u-(q+\alpha)\th u\big]=0,
\end{equation}
where $\alpha,\beta$ are constants, and its Lax pair reads (cf.
\cite{Bridgman-Lax})
\begin{subequations}\label{NQC-lax}
\begin{align}\label{NQC-lax-t}
&\t \phi=\gamma_1\left(
\begin{array}{cc}
(p-\alpha)(p-\beta)u-(p^2-r^2)\t u&-(r-\alpha)(r-\beta)u\t u\\
(r+\alpha)(r+\beta)&-(p+\alpha)(p+\beta)\t u+(p^2-r^2)u
\end{array} \right)\phi,\\
&\h \phi=\gamma_2\left(
\begin{array}{cc}
(q-\alpha)(q-\beta)u-(q^2-r^2)\h u&-(r-\alpha)(r-\beta)u\h u\\
(r+\alpha)(r+\beta)&-(q+\alpha)(q+\beta)\h u+(q^2-r^2)u
\end{array} \right)\phi,
\end{align}
\end{subequations}
where $\phi=(\phi_1,\phi_2)^T$, $\gamma_1$ is either
\begin{subequations}
\begin{align}\label{NQC-p1}
\gamma_1=\frac{1} {\sqrt{\big[(\beta-p)u+(\alpha+p)\t
u\big] \big[(\alpha-p)u+(\beta+p)\t u\big]}},\\
\label{NQC-p2}
\text{or}~~\gamma_1=\frac{1}{(\alpha-p)u+(\beta+p)\t
u},~~\text{or}~~\gamma_1=\frac{1}{(\beta-p)u+(\alpha+p)\t u},
\end{align}
\end{subequations}
and  $\gamma_2$ follows from the above $\gamma_1$'s by replacing $(p,\t ~ \,)$ by $(q,\h ~ \,)$.

Eliminating $\phi_1$ from \eqref{NQC-lax-t} one finds
\begin{subequations}
\begin{align}\label{NQC-Ri-t}
A\t{\t{\phi}}_2 + B\t\phi_2 + \varepsilon C\phi_2=0,
\end{align}
where $\varepsilon=p^2-r^2$,
\begin{align}
&A=\frac{1}{\t\gamma_1},~~B=(p+\alpha)(p+\beta)\t{\t{u}}-(p-\alpha)(p-\beta)u,\\
&C=\gamma_1 \big[(\alpha^2+\beta^2-2p^2)u\t u +
(p-\alpha)(p-\beta)u^2 + (p+\alpha)(p+\beta)\t u^2\big].
\end{align}
\end{subequations}
\eqref{NQC-Ri-t} yields a discrete Riccati equation
\begin{align}\label{NQC-Ri}
A\t \theta\theta + B\theta + \varepsilon C=0,
\end{align}
with $\theta=\t\phi_2/\phi_2$, which is then solved by
\begin{subequations}\label{theta}
\begin{align}
\theta=\rho\varepsilon(1+\sum_{j=1}^\infty\theta_j\varepsilon^j),
\end{align}
with
\begin{align}
\rho=-\frac{C}{B},~~~
\theta_{j+1}=-\frac{A\t\rho}{B}\sum_{i=0}^{j}\t\theta_i\theta_{j-i},\quad(\theta_0=1),~~ j=0,1,2,\cdots.
\label{NQC-theta}
\end{align}
\end{subequations}
Next, going back to the Lax pair \eqref{NQC-lax} we can easily find
\begin{subequations}
\begin{align}
&\theta=\frac{\t\phi_2}{\phi_2}=\gamma_1\big[(r+\alpha)(r+\beta)\frac{\phi_1}{\phi_2}-(p+\alpha)(p+\beta)\t
u+(p^2-r^2)u\big],\\
&\eta=\frac{\h\phi_2}{\phi_2}=\gamma_2\big[(r+\alpha)(r+\beta)\frac{\phi_1}{\phi_2}-(q+\alpha)(q+\beta)\h
u+(q^2-r^2)u\big],
\end{align}
\end{subequations}
from which eliminating $\phi_1/\phi_2$ we reach to the relation
\begin{subequations}
\begin{align}\label{eta}
\eta=\omega(1+\sigma\theta),
\end{align}
with
\begin{align}
&\omega=\gamma_2\big[(p+\alpha)(p+\beta)\t u-(q+\alpha)(q+\beta)\h
u+(q^2-p^2)u\big],\label{NQC-omega}\\
&\sigma=\frac{1}{\gamma_1\big[(p+\alpha)(p+\beta)\t
u-(q+\alpha)(q+\beta)\h u+(q^2-p^2)u\big]}\label{NQC-sigma}.
\end{align}
\end{subequations}
Meanwhile, due to $
\theta=\t{\phi}_2/\phi_2,~~\eta=\h{\phi}_2/\phi_2, $ we get
\begin{equation}\label{cl-def}
\Delta_m\ln\theta=\Delta_n\ln\eta,
\end{equation}
which provides a formal conservation law for the NQC equation.
Finally, what we need is to insert the explicit form \eqref{theta}
of $\theta$ into \eqref{cl-def} and then expand it in terms of
$\varepsilon$. The coefficient of each power of $\varepsilon$
provides a conservation law for the NQC equation, which is expressed
through (cf. \cite{zcs-cls-ABS})
\begin{subequations}\label{CLs-inf}
\begin{align}
& \Delta_m \ln \rho=\Delta_n \ln \omega,\\
& \Delta_m\, h_s(\boldsymbol\theta)=\Delta_n\,
h_s(\sigma\rho\underline{\boldsymbol\theta}),~~ (s=1,2,3,\cdots),
\end{align}
where
\begin{equation}
\boldsymbol\theta=(\mathbf{\theta}_1,\theta_2,\cdots),~~~
\underline{\boldsymbol\theta}=(1,\theta_1,\theta_2,\cdots),~~\text{and}~~
\sigma\rho\underline{\boldsymbol\theta}=(\sigma\rho,\sigma\rho\theta_1,\sigma\rho\theta_2,\cdots),
\end{equation}
with $\rho$, $\omega$, $\sigma$ and $\{\theta_j\}$ given by
\eqref{NQC-theta}, \eqref{NQC-omega}, \eqref{NQC-sigma} and
\eqref{NQC-theta}.
\end{subequations}
$\{h_s(\bft)\}$ are polynomials defined as the following\cite{zcs-cls-ABS}.

\begin{prop}\label{P:2}
The following expansion holds,
\begin{subequations}\label{exp-t}
\begin{equation}
\ln\biggl(1+\sum_{i=1}^{\infty}t_i k^i
\biggr)=\sum_{j=1}^{\infty}h_j(\bft)k^{j}, \label{ht}
\end{equation}
where
\begin{equation}
h_j(\bft)=\sum_{||\balpha||=j}(-1)^{|\balpha|-1}(|\balpha|-1)!\frac{\bft^{\balpha}}{\balpha
!}, \label{htj}
\end{equation}
and
\begin{align}
& \mathbf{t}=(t_1,t_2,\cdots),~~\balpha=(\alpha_1,\alpha_2,\cdots),~~\alpha_i\in\{0,1,2,\cdots\}, \\
& \bft^{\balpha}=\prod_{i=1}^{\infty}t_i^{\alpha_i},~~
 {\balpha}!=\prod_{i=1}^{\infty}(\alpha_i !),~~
|\balpha|=\sum_{i=1}^{\infty}\alpha_i,~~
||\balpha||=\sum^{\infty}_{i=1} i\alpha_i.
\end{align}
\end{subequations}
The first few of $\{h_j(\bft)\}$ are
\begin{subequations}\label{ht1-4}
\begin{align}
& h_1(\bft)=t_1,\\
& h_2(\bft)=-\frac{1}{2}t_1^2+t_2, \\
& h_3(\bft)=\frac{1}{3}t_1^3-t_1t_2+t_3,\\
& h_4(\bft)=-\frac{1}{4}t_1^4+t_1^2t_2-t_1t_3-\frac{1}{2}t_2^2+t_4.
\end{align}
\end{subequations}
\end{prop}

\subsection{The DBSQ equation}

Now let us look at the DBSQ equation\cite{TN-DBSQ}
\begin{equation}
\t z-x\t x+y=0,~~\h z-x\h x+y=0,~~(\h x-\t x)(z-x\th x+\th y)-p+q=0.
\label{DBSQ}
\end{equation}
Its Lax pair reads
\begin{subequations}
\label{lax-db}
\begin{align}
&\t\phi=\left(
\begin{array}{ccc}
 -\t x&1&0\\-\t y&0&1\\p-r-x\t
y+\t x z&-z&x
\end{array}
\right)\phi,\label{lax-db1}\\
&\h\phi=\left(
\begin{array}{ccc}
 -\h x&1&0\\-\h y&0&1\\q-r-x\h
y+\h x z&-z&x
\end{array}
\right)\phi,\label{lax-db2}
\end{align}
\end{subequations}
where $\phi=(\phi_1,\phi_2,\phi_3)^T$. From \eqref{lax-db1} we can
eliminate $\phi_2,\phi_3$ and get \[\t{\t{\t\phi}}_1 + (\t{\t{\t
x}}-x)\t{\t{\phi}}_1 + (\t{\t{y}}+z-x\t{\t{x}})\t\phi_1 +
\varepsilon\phi_1=0,\] where $\varepsilon=r-p,$
 and then a discrete Riccati equation
\begin{equation}
\t{\t\theta}\t\theta\theta + (\t{\t{\t x}}-x)\t\theta\theta +
(\t{\t{y}}+z-x\t{\t{x}})\theta + \varepsilon=0, \label{riccati-db}
\end{equation}
with $\theta=\t\phi_1/\phi_1$. This is a third-order equation and
solved by
\begin{subequations}\label{DBSQ-theta}
\begin{align}
\theta=\rho\varepsilon\Bigl(1+\sum_{j=1}^\infty\theta_j\varepsilon^j\Bigr),
\end{align}
with
\begin{align}
&\rho=-\frac{1}{\t{\t{y}}+z-x\t{\t{x}}},\label{rho-db}\\
&\theta_1=-\frac{\t \rho (\t{\t{\t x}}-x)}{\t{\t{y}}+z-x\t{\t{x}}},\label{theta1-db}\\
&\theta_{j+2}=-\frac{\t
\rho}{\t{\t{y}}+z-x\t{\t{x}}}\Bigl[\t{\t{\rho}}\sum_{i=0}^j\sum_{k=0}^{j-i}\t{\t{\theta}}_i\t{\theta}_k\theta_{j-i-k}
+ (\t{\t{\t x}}-x)\sum_{i=0}^{j+1}\t \theta_i\theta_{j+1-i}
\Bigr],~(\theta_0=1)\label{thetai-db},
\end{align}
\end{subequations}
for $j=0,1,2,\cdots.$
Meanwhile, from the Lax pair \eqref{lax-db} we have
\begin{align*}
&\theta=\frac{\t\phi_1}{\phi_1}=-\t x + \frac{\phi_2}{\phi_1},\\
&\eta=\frac{\h\phi_1}{\phi_1}=-\h x + \frac{\phi_2}{\phi_1},
\end{align*}
which yields
\begin{subequations}
\begin{align}\label{eta-db}
\eta=\omega(1+\sigma\theta),
\end{align}
with
\begin{align}
&\omega=\t x-\h x,\label{omega-db}\\
&\sigma=\frac{1}{\t x-\h x}\label{sigma-db}.
\end{align}
\end{subequations}
Next, from the formal conservation law
$\Delta_m\ln\theta=\Delta_n\ln\eta$, we get infinitely many
conservation laws
\begin{subequations}\label{CLs-inf}
\begin{align}
& \Delta_m \ln \rho=\Delta_n \ln \omega,\\
& \Delta_m\, h_s(\boldsymbol\theta)=\Delta_n\,
h_s(\sigma\rho\underline{\boldsymbol\theta}),~~ (s=1,2,3,\cdots),
\end{align}
where
\begin{equation}
\boldsymbol\theta=(\mathbf{\theta}_1,\theta_2,\cdots),~~~
\underline{\boldsymbol\theta}=(1,\theta_1,\theta_2,\cdots),~~\text{and}~~
\sigma\rho\underline{\boldsymbol\theta}=(\sigma\rho,\sigma\rho\theta_1,\sigma\rho\theta_2,\cdots),
\end{equation}
with $\rho$, $\omega$, $\sigma$ and $\{\theta_j\}$ given by
\eqref{rho-db}, \eqref{omega-db}, \eqref{sigma-db},
\eqref{theta1-db} and \eqref{thetai-db}.
\end{subequations}
$\{h_s(\bft)\}$ are polynomials defined in  Propositon \ref{P:2}.

\section{Conservation laws of some multi-component lattice
equations}\label{sec3}

\subsection{Generic description}
We first list all multi-component lattice equations involved in this
part.
\begin{align*}
&\t y-\h y-y[(\t x-\h x)y+p-q]=0,~~\t x-\h x+\th x[(\t x-\h
x)y+p-q]=0,\tag*{lNLS}
\\
&\th{x}(p\t y-q\h y)-y(p\h x-q\t x)=0,~~ x\th{y}(p\t y-q\h y)-y(p\t x\h y-q\h x\t y)=0,\tag*{mDBSQ}\\
&\th{x}-\frac{\h x\t z-\t x\h z}{\t z-\h
z}=0,~~\th{z}+z\th{x}-\frac{z(p\h{z}-q\t z)}{\t z-\h z}=0,\tag{C-2.1}\\
&\th{x}-\frac{\h x\t z-\t x\h z}{\t z-\h
z}=0,~~\th{z}+d\frac{z}{x}-\frac{z}{x}\frac{p\t x\h z-q\h x\t z}{\t
z-\h z}=0,\tag{C-2.2}
\\
&\t xz-\t y-x=0,~~\h xz-\h y-x=0,~~\th{z}-\frac{y}{x}-\frac{1}{x}\frac{p\t x-q\h x}{\t z-\h z}=0,\tag{A-2}\\
&x\t x-\t y-z=0,~~x\h x-\h y-z=0,~~\th{z}+y-d(\th{x}-x)-x\th{x}-\frac{p-q}{\t x-\h x}=0,\tag{B-2}\\
&z\t y-\t x+x=0,~~z\h y-\h x+x=0,~~\th{z}-\frac{d_2x+d_1}{y}-\frac{z}{y}\frac{p\t y\h z-q\h y\t z}{\t z-\h z}=0,\tag{C-3}\\
&z\t y-\t x+x=0,~~z\h y-\h
x+x=0,~~\th{z}-\frac{x\th{x}+d}{y}-\frac{z}{y}\frac{p\t y\h z-q\h
y\t z}{\t z-\h z}=0,\tag{C-4}\\
&\t z-y\t x-z=0,~~\h z-y\h x-z=0,~~x\th y(\t y-\h y)-y(p\t x\h y-q\h
x\t y)=0,\tag*{SDBSQ}\\
&\left\{\begin{array}{l}
\th{y}(p-q+\h x-\t x)-(p-1)\h y+(q-1)\t y=0,\\
\t y\h y(p-q-\h z+\t z)-(p-1)y\h y+(q-1)y\t y=0,\\
y(p+q-z-\th{x})(p-q+\h x-\t x)-(p^2+p+1)\t y+(q^2+q+1)\h y=0.
\end{array}
\right.
\tag*{Toda-mDBSQ}
\end{align*}
All these equations are of multi-component, defined on an elementary
quadrilateral, and multi-dimensionally consistent in terms of the
vector variable $u=(x,y,z)$. For some two-component equations $z$ or
$y$ is absent. Among these equations, lNLS stands for lattice
nonlinear Schr\"odinger equation given in \cite{Mikhailov}, mDBSQ
stands for modified discrete Boussinesq equation given in
\cite{Walker-2001}, (C-2.1), (C-2.2), (A-2), (B-2), (C-3) and
(C-4) are the lattice equations of Boussinesq type found in
\cite{Hietarinta-Bous}, SDBSQ stands for Schwarzian discrete
Boussinesq equation  given in \cite{Nijhoff-Sch-eq}, and Toda-mDBSQ
stands for Toda-modified discrete Boussinesq equation given in
\cite{Nijhoff-LGD}. Obviously, the DBSQ equation can be obtained
from (B-2) by setting $d=0$ and switching
$(x,y,z,p,q)\rightarrow(x,z,y,q,p)$, and the SDBSQ equation can be
obtained from (C-3) by setting $d_1=d_2=0$ and switching
$(x,y,z)\rightarrow(z,x,y)$. The Lax pairs of all these lattice
equations are listed in Ref.\cite{Bridgman-Lax}, while we list  them
in Appendix \ref{A:A}.

It is possible to describe a unified approach to derive infinitely many conservation laws
for all the above mentioned  multi-component lattice equations.
Their Lax pairs are of the following form
\begin{subequations}
\begin{align}
\t{\phi}=N_1\phi, \label{lax-1}\\
\h{\phi}=N_2\phi,
\end{align}
\end{subequations}
where $N_1$ and $N_2$ are $N\times N$ matrices  and $\phi=(\phi_1,\phi_2,\cdots,\phi_N)^T$.
There is some certain $\phi_{i_0}$ such that one can from \eqref{lax-1} eliminate other $\phi_j$'s
and get a scalar form spectral problem in terms of $\phi_{i_0}$, say, the following
\begin{equation}
A\,\t {\t {\t \phi}}_{i_0}+B\,\t{\t \phi}_{i_0} + (\varepsilon C+D)\t\phi_{i_0}+\varepsilon G \phi_{i_0}=0,
\end{equation}
where $A,B,C,D,G$ are functions of $(E_n^ju,p)$, and $\varepsilon$
is a constant related to $p$ and $r$. From this we reach to a
discrete Riccati equation
\begin{equation}\label{Riccati}
A\t{\t{\theta}}\t{\theta}\theta+B\t \theta\theta+ (\varepsilon C+D)\theta+\varepsilon G=0,
\end{equation}
with
\begin{equation}
\theta=\frac{\t\phi_{i_0}}{\phi_{i_0}}.
\end{equation}
As for solutions to \eqref{Riccati} we have
\begin{prop}
The discrete Riccati equation \eqref{Riccati} is solved by
\begin{subequations}\label{multi-l-theta}
\begin{align}
\theta=\rho\varepsilon\Bigl(1+\sum_{j=1}^\infty\theta_j\varepsilon^j\Bigr),
\end{align}
with
\begin{align}
&\rho=-\frac{G}{D},\\
&\theta_1=-\frac{1}{D} (B\t \rho + C),\\
&\theta_{j+2}=-\frac{1}{D}\Bigl(A\t
\rho\t{\t{\rho}}\sum_{i=0}^j\sum_{k=0}^{j-i}\t{\t{\theta}}_i\t{\theta}_k\theta_{j-i-k}+B\t
\rho \sum_{i=0}^{j+1}\t
\theta_i\theta_{j+1-i}+C\theta_{j+1}\Bigr),~(\theta_0=1),
\end{align}
\end{subequations}
for $j=0,1,2,\cdots.$
\end{prop}

Next, the following relation is also available (recalling \eqref{eta-db}),
\begin{align}\label{omega-sigma}
\eta=\frac{\h\phi_{i_0}}{\phi_{i_0}}=\omega(1+\sigma\theta),
\end{align}
where  $\omega$ and $\sigma$ are functions of $(u,\t u,\h u,p,q)$
related to considered equations and they satisfy
\begin{equation}
\omega(u,\t u,\h u,p,q)=-\frac{1}{\sigma(u,\h u,\t u,q,p)}.
\label{sym}
\end{equation}
Then, the infinitely many conservation laws can be described as following (cf.\cite{zcs-cls-ABS}).

\begin{prop}\label{P:3}
From the formal conservation law
\begin{equation}\label{multi-l-CLs}
\Delta_m\ln\theta=\Delta_n\ln\eta,
\end{equation}
one has
\begin{subequations}
\begin{align}
& \Delta_m \ln \rho=\Delta_n \ln \omega,\label{cls-1}\\
& \Delta_m\, h_s(\boldsymbol\theta)=\Delta_n\,
h_s(\sigma\rho\underline{\boldsymbol\theta}),~~ (s=1,2,3,\cdots),
\end{align}
where
\begin{equation}
\boldsymbol\theta=(\mathbf{\theta}_1,\theta_2,\cdots),~~~
\underline{\boldsymbol\theta}=(1,\theta_1,\theta_2,\cdots),
\end{equation}
with $\rho,\{\theta_i\},\omega$ and $\sigma$ given by
\eqref{multi-l-theta} and \eqref{omega-sigma} and $h_s(\bf t)$
defined in Proposition \ref{P:2}.
\end{subequations}
The first few conservation laws are
\begin{subequations}
\begin{align}
\Delta_m \ln\Big(-\frac{G}{D}\Big)&=\Delta_n \ln \omega,\\
\Delta_m\frac{C\t D-B\t G}{D\t D}&=\Delta_n\frac{G\sigma}{D},\\
\Delta_m\Bigl[\frac{(B\t G-C\t D)^2}{2D^2\t D^2}+\frac{B\t G(\t
B\t{\t G}-\t C\t{\t D})}{D\t D^2\t{\t D}} - \frac{A\t G\t{\t G}}{D\t
D\t{\t{D}}}\Bigr]&=\Delta_n\frac{G\sigma}{2D^2\t D}\Bigl[2(C\t D-B\t
G)-\t DG\sigma\Bigr].
\end{align}
\end{subequations}
\end{prop}


\subsection{Main results}
 We find  each multi-component lattice system we list out in our paper
 falls in the above frame and therefore they can share those formulae of conservation laws
 with concrete $\{A,B,C,D,G,\omega,\sigma\}$ where in some cases $A$ can also be scaled to 1. In the following we skip details and list out $A,B,C,D,G,\omega$ and
$\sigma$ for each equation.

\begin{prop}\label{P:4}
For lNLS equation, $i_0=1$,
\begin{subequations}
\begin{align}
&A=0,~B=\frac{1}{\t{\t{x}}},~C=\frac{1}{\t
x},~D=\frac{1+\t{\t{x}}y}{\t{\t{x}}},~G=\frac{1}{\t
x},~\omega=\frac{\h x-\t x}{\t x},~\sigma=\frac{\h x}{\h x-\t x}.
\end{align}
\end{subequations}
For mDBSQ equation, $i_0=3$,
\begin{subequations}
\begin{align}
&A=\frac{1}{\t\gamma_1\t{\t{\gamma}}_1\t
y\t{\t{y}}\t{\t{\t{y}}}},~B=-\frac{p[\t{\t{\t{y}}}(\t xy+\t{\t{x}}\t
y)+\t{\t{\t{x}}}y\t{\t{y}}]}{\t\gamma_1\t{\t{x}}y\t
y\t{\t{y}}\t{\t{\t{y}}}},~C=0,~G=\gamma_1,\\
&D=\frac{p^2(x\t{\t{y}}+\t xy+\t{\t{x}}\t y)}{\t
xy\t{\t{y}}},~\omega=\frac{\gamma_2y(q\h x\t y-p\t x\h y)}{x\t
y},~\sigma=\frac{x\h y}{\gamma_1y(q\h x\t y-p\t x\h y)}.
\end{align}
\end{subequations}
For (C-2.1) equation, $i_0=3$,
\begin{subequations}
\begin{align}
&A=\frac{1}{\t\gamma_1\t{\t{\gamma}}_1\t
z\t{\t{z}}},~B=\frac{1}{\t\gamma_1z\t
z\t{\t{z}}}[z\t{\t{\t{z}}}+z\t z(p+\t{\t x})+\t z\t{\t z}],~C=1,\\
&D=\t{\t{x}}+\frac{\t{\t{z}}}{z}+p,~G=\gamma_1\t
z,~\omega=\gamma_2(\t z-\h z),~\sigma=\frac{1}{\gamma_1(\t z-\h z)}.
\end{align}
\end{subequations}
For (C-2.2) equation, $i_0=3$,
\begin{subequations}
\begin{align}
&A=\frac{\t x}{\t\gamma_1\t{\t{\gamma}}_1\t
z\t{\t{z}}},~B=\frac{1}{\t\gamma_1z\t z\t{\t{z}}}[x\t z\t{\t{z}}+\t
x
z\t{\t{\t{z}}}+(d+p\t{\t{x}})z\t z],~C=\t x,\\
&D=\frac{x\t{\t{z}}}{z}+p\t{\t{x}}+d,~G=\gamma_1\t x\t
z,~\omega=\gamma_2(\t z-\h z),~\sigma=\frac{1}{\gamma_1(\t z-\h z)}.
\end{align}
\end{subequations}
For (A-2) equation, $i_0=3$,
\begin{subequations}
\begin{align}
&A=\frac{\t
x}{\t\gamma_1\t{\t{\gamma_1}}\t{\t{z}}},~B=\frac{1}{\t\gamma_1}(-x+\t
x\t{\t{\t{z}}}-\t y),~C=0,~G=\gamma_1 \t xz\t z,\\
&D=\t z(p\t{\t{x}}-x\t{\t{z}}+y),~\omega=\gamma_2z(\t z-\h
z),~\sigma=\frac{1}{\gamma_1z(\t z-\h z)}.
\end{align}
\end{subequations}
For (B-2) equation, $i_0=3$,
\begin{subequations}
\begin{align}
&A=\frac{1}{\t{\gamma_1}\t{\t{\gamma_1}}\t
x\t{\t{x}}},~B=\frac{1}{\t \gamma_1\t
x}(x-\t{\t{\t{x}}}+d),~C=0,~G=\gamma_1x,\\
&D=d(x-\t{\t{x}})-x\t{\t{x}}+y+\t{\t{z}},~\omega=\gamma_2x(\h x-\t
x),~\sigma=\frac{1}{\gamma_1x(\h x-\t x)}.
\end{align}
\end{subequations}
For (C-3) equation, $i_0=3$,
\begin{subequations}
\begin{align}
&A=\frac{\t y}{\t\gamma_1\t{\t{\gamma}}_1\t
z\t{\t{z}}},~B=\frac{1}{\t\gamma_1\t z\t{\t{z}}}(y\t{\t{z}}+\t
y\t{\t{\t{z}}}+p\t{\t{y}}\t
z-d_2\t x-d_1),~C=0,\\
&D=\frac{1}{\t z}[y\t{\t{z}}+p(\t y z+\t{\t{y}}\t
z)-d_2x-d_1],~G=\gamma_1(\t x-x),~\omega=\gamma_2(\t z-\h
z),~\sigma=\frac{1}{\gamma_1(\t z-\h z)}.
\end{align}
\end{subequations}
For (C-4) equation, $i_0=2$,
\begin{subequations}
\begin{align}
&A=\frac{1}{\t\gamma_1\t{\t{\gamma}}_1(\t{\t{\t{x}}}-\t{\t
x})},~B=\frac{\t{\t{z}}}{\t\gamma_1(\t{\t{\t{x}}}-\t{\t{x}})}+
\frac{-x\t x+p\t yz+y\t{\t{z}}-d}{y\t\gamma_1(\t{\t{x}}-\t x)},~C=0,\\
&D=\frac{\t z[-x\t{\t{x}}+y\t{\t{z}}+p(\t yz+\t{\t{y}}\t
z)-d]}{y(\t{\t{x}}-\t x)},~G=\frac{\gamma_1z\t
z}{y},~\omega=\frac{\gamma_2z(\h x-\t x)}{\t x-x},~\sigma=\frac{\h
x-x}{\gamma_1z(\h x-\t x)}.
\end{align}
\end{subequations}
For SDBSQ equation, $i_0=3$,
\begin{subequations}
\begin{align}
&A=\frac{\t x}{\t\gamma_1\t{\t{\gamma}}_1\t{\t{y}}},
~B=-\frac{1}{\t\gamma_1\t{\t{y}}}(x\t{\t{y}}+\t{x}\t{\t{\t{y}}}+p\t{\t{x}}\t{y}),
~C=0,\\
&D=p(\t xy+\t{\t{x}}\t y )+x\t{\t{y}},~G=\gamma_1\t xy\t
y,~\omega_2=\gamma_2(\h y-\t y),~\sigma=\frac{1}{\gamma_1(\h y-\t
y)}.
\end{align}
\end{subequations}
For Toda-mDBSQ equation, $i_0=2$,
\begin{subequations}
\begin{align}
&A=\frac{1}{\t\gamma_1\t{\t{\gamma_1}}\t{\t{\t{y}}}},
~B=\frac{1-p}{\t\gamma_1\t{\t{\t{y}}}}+\frac{\t{\t{x}}+z-2p}{\t\gamma_1\t{\t{y}}},~C=0,
~D=\frac{p-1}{\t{\t{y}}}(-\t{\t{x}}-z+2p)+\frac{p^2+p+1}{y},\\
&G=\frac{\gamma_1}{y},~\omega=\frac{\gamma_2[(q-1)\t y-(p-1)\h
y]}{\t y},~\sigma=\frac{\h y}{\gamma_1[(q-1)\t y-(p-1)\h y]}.
\end{align}
\end{subequations}
For each equation, the function $\gamma_j$ is defined in Appendix A.
\end{prop}
For each equation, from Proposition 2 and Proposition 4 we can find
that $\rho$ is related to $\gamma_1$
 and $\omega$ is related to $\gamma_2$ while $\{\theta_j\}$ and $\sigma\rho$ are independent of $\gamma_1$ and $\gamma_2$, thus by Proposition 3 all conservation
laws except the first one \eqref{cls-1} are independent of $\gamma_1$ and $\gamma_2$.

\section{Conclusion}
We have shown some examples of deriving infinitely many conservation
laws from Lax pairs for some lattice equations, particularly for
multi-component discrete systems. These systems are all integrable
in the sense of multi-dimensional consistency. Such integrability is
used to construct Lax pairs. In \cite{NQC-cls-2006} three-point
conservation laws were found via direct approach. Here the simplest
nontrivial conservation law of the NQC equation is a four-point one
(see Appendix \ref{A:B}). However, the approach using Lax pairs
looks quite natural and can provide infinitely many conservation
laws. And more important, it works for most of known
multi-dimensionally consistent systems, including one-component and
multi-component discrete systems. We also note that if we conduct
the same procedure starting from $(q,\widehat{~}\,)$ part of Lax
pairs, we only need to switch $(p,\widetilde{~}\,)$ and
$(q,\widehat{~}\,)$ in the present results and this is guaranteed by
the symmetric property \eqref{sym}.

\subsection*{Acknowledgments}
The authors are grateful to Professor Hereman for kindly sending Ref. \cite{Bridgman-Lax}.
This project is supported by the NSF of China
(No. 11071157), SRF of the DPHE of China (No. 20113108110002) and
Shanghai Leading Academic Discipline Project (No. J50101).

\begin{appendix}

\section{Lax pairs of lattice equations listed in Sec.\ref{sec3} (cf.\cite{Bridgman-Lax})}\label{A:A}
For each equation we only list out the matrix $N_1$ in the Lax pair.
Matrix $N_2$ follows from $N_1$ by switching $(1,p,\t ~ ) \rightarrow (2,q,\h ~)$.

For the lNLS equation
\begin{equation*}
\begin{aligned}
N_1=\gamma_1\left(\begin{array}{cc} -1&\t x\\y&r-p-y\t x
\end{array}
\right),
\end{aligned}~
{\footnotesize\begin{gathered} \text{with}~\gamma_1=1.
\end{gathered}}
\end{equation*}
For the mDBSQ equation
\begin{equation*}
\begin{aligned}
N_1=\gamma_1\left(\begin{array}{ccc}p\t y&0&-r\\
-r\t xy&py&0\\
0&-\frac{ry\t y}{x}&\frac{p\t xy}{x}
\end{array}
\right),
\end{aligned}~
{\footnotesize\begin{gathered}
\text{with}~\gamma_1=\sqrt[3]{\frac{x}{\t xy^2\t
y}},~\text{or}~\gamma_1=\frac{1}{y},~\text{or}~\gamma_1=\frac{1}{\t
y}.
\end{gathered}}
\end{equation*}
For (C-2.1) equation
\begin{equation*}
\begin{aligned}
N_1=\gamma_1\left(\begin{array}{ccc} -\t z&\t x&0\\z\t z&-z(p+\t
x)&rz\t z\\0&1&-\t z
\end{array}
\right),
\end{aligned}~
{\footnotesize\begin{gathered}
\text{with}~\gamma_1=\frac{1}{\sqrt[3]{z\t
z^2}},~\text{or}~\gamma_1=\frac{1}{z},~\text{or}~\frac{1}{\t z}.
\end{gathered}}
\end{equation*}
For (C-2.2) equation
\begin{equation*}
\begin{aligned}
N_1=\gamma_1\left(\begin{array}{ccc} -\t z&\t x&0\\
\frac{rz\t z}{x}& -\frac{z}{x}(d+p\t x)&\frac{dz\t z}{x}\\
0&1&-\t z
\end{array}
\right),
\end{aligned}~
{\footnotesize\begin{gathered}
\text{with}~\gamma_1=\sqrt[3]{\frac{x}{\t xz\t
z^2}},~\text{or}~\gamma_1=\frac{1}{z},~\text{or}~\frac{1}{\t z}.
\end{gathered}}
\end{equation*}
For (A-2) equation
\begin{equation*}
\begin{aligned}
N_1=\gamma_1\left(
\begin{array}{ccc}
\frac{yz}{x}&\frac{r}{x}&\frac{rx-p\t xz-yz\t z}{x}\\
-\t x z&\t z&x\t z\\
z&0&-z\t z
\end{array}
\right),
\end{aligned}~
{\footnotesize\begin{gathered}
\text{with}~\gamma_1=\sqrt[3]{\frac{x}{\t xz^2\t
z}},~\text{or}~\gamma_1=\frac{1}{z},~\text{or}~\gamma_1=\frac{1}{\t
z}.
\end{gathered}}
\end{equation*}
For (B-2) equation
\begin{equation*}
\begin{gathered} N_1=\gamma_1\left(
\begin{array}{ccc}
-(d x+x^2)&d x+y&k_1\\
-x\t x&\t z&z\t z\\
0&-1&x\t x-z
\end{array}
\right),
\end{gathered}~
{\footnotesize \begin{gathered}
\text{where}~k_1=(z-x\t x)(dx+y)\\+\t z(dx+x^2)+x(p-r),\\
\text{with}~\gamma_1=\frac{1}{\sqrt[3]{x^2\t
x}},~\text{or}~\gamma_1=\frac{1}{x},~\text{or}~\gamma_1=\frac{1}{\t
x}.
\end{gathered}}
\end{equation*}
For (C-3) equation
\begin{equation*}
\begin{aligned}
N_1=\gamma_1\left(\begin{array}{ccc} \frac{d_1+d_2x-p\t
yz}{y}&\frac{rz\t z}{y}&-\frac{d_1\t z+d_2x\t z}{y}\\
0&-z&\t x-x\\
1&0&-\t z
\end{array}
\right),
\end{aligned}~
{\footnotesize\begin{gathered}
\text{with}~\gamma_1=\sqrt[3]{\frac{y}{\t yz^2\t
z}},~\text{or}~\gamma_1=\frac{1}{z},~\text{or}~\gamma_1=\frac{1}{\t
z}.
\end{gathered}}
\end{equation*}
For (C-4) equation
\begin{equation*}
\begin{aligned}
N_1=\gamma_1\left(\begin{array}{ccc} \frac{d+x\t x-p\t
yz}{y}&\frac{(r-x)z\t z}{y}&-\frac{(d+x^2)\t z}{y}\\
0&-z&\t x-x\\
1&0&-\t z
\end{array}
\right),
\end{aligned}~
{\footnotesize\begin{gathered}
\text{with}~\gamma_1=\sqrt[3]{\frac{y}{\t yz^2\t
z}},~\text{or}~\gamma_1=\frac{1}{z},~\text{or}~\gamma_1=\frac{1}{\t
z}.
\end{gathered}}
\end{equation*}
For the SDBSQ equation
\begin{equation*}
\begin{aligned}
N_1=\gamma_1\left(\begin{array}{ccc} \frac{py\t x}{x}&-\frac{r\t
y}{x}&\frac{rz\t y}{x}\\
-\t z&\t y&0\\
-1&0&\t y
\end{array}
\right),
\end{aligned}~
{\footnotesize\begin{gathered}
\text{with}~\gamma_1=\sqrt[3]{\frac{x}{\t y^2(\t
z-z)}},~\text{or}~\gamma_1=\frac{1}{y},~\text{or}~\gamma_1=\frac{1}{\t
y}.
\end{gathered}}
\end{equation*}
For the Toda-mDBSQ equation
\begin{equation*}
\begin{aligned}
N_1=\gamma_1\left(\begin{array}{ccc}r+p-z&\frac{1+r+r^2}{y}&k_1\\
0&p-1&(1-r)\t y\\
1&0&p-r-\t x
\end{array}
\right),
\end{aligned}~
{\footnotesize\begin{gathered} \text{where}~k_1=(p^2-r^2)-\t
x(p+r)\\+z(r-p+\t x)
-\frac{\t y}{y}(p^2+p+1),\\
\text{with}~\gamma_1=\sqrt[3]{\frac{y}{\t
y}},~\text{or}~\gamma_1=1.
\end{gathered}}
\end{equation*}

\section{First few conservation laws of some lattice equations}\label{A:B}

For the NQC equation, the first two conservation laws are
\begin{subequations}
\begin{align}
&\Delta_m\ln\frac{\gamma_1 \big[(\alpha^2+\beta^2-2p^2)u\t u +
P_-u^2 + P_+\t
u^2\big]}{P_-u-P_+\t{\t{u}}}=\Delta_n\ln\gamma_2\big[P_+\t
u-Q_+\h u+(q^2-p^2)u\big],\\
&\Delta_m\frac{(\alpha^2+\beta^2-2p^2)\t u\t{\t u} + P_-\t u^2 +
P_+\t{\t u}^2}{(P_-u-P_+\t{\t u})(P_-\t u-P_+\t{\t{\t u}})}=
\Delta_n\frac{(\alpha^2+\beta^2-2p^2)u\t u + P_-u^2 + P_+\t
u^2}{[P_+\t u-Q_+\h u+(q^2-p^2)u](P_-u-P_+\t{\t{u}})},
\end{align}
where
$P_+=(p+\alpha)(p+\beta),~P_-=(p-\alpha)(p-\beta),~Q_+=(q+\alpha)(q+\beta).$
\end{subequations}
For the DBSQ equation, the first two conservation laws are
\begin{subequations}
\begin{align}
\Delta_m\ln\frac{1}{x\t{\t x}-\t{\t y}-z}=&\Delta_n\ln(\t x-\h x),\\
\Delta_m\frac{-x+\t{\t{\t x}}}{(x\t{\t x}-\t{\t y}-z)(\t x\t{\t{\t
x}}-\t{\t{\t y}}-\t z)}=&\Delta_n\frac{1}{(\t x-\h x)(x\t{\t
x}-\t{\t y}-z)}.
\end{align}
\end{subequations}
For the lNLS equation, the first two conservation laws are
\begin{subequations}
\begin{align}
\Delta_m\ln\frac{-\t{\t{x}}}{\t x(1+\t{\t{x}}y)}=&
\Delta_n\ln\frac{\h x-\t x}{\t x},\\
\Delta_m\frac{\t x\t{\t{\t x}}-\t{\t x}^2(1+\t{\t{\t x}}\t y)}{\t
x\t{\t x}(1+\t{\t x}y)(1+\t{\t{\t x}}\t y)}=&\Delta_n\frac{\h x\t{\t
x}}{\t x(1+\t{\t x}y)(\t x-\h x)}.
\end{align}
\end{subequations}
For the SDBSQ equation, the first two conservation laws are
\begin{subequations}
\begin{align}
\Delta_m\ln\frac{-\gamma_1\t xy\t y}{p(\t xy+\t{\t x}\t
y)+x\t{\t y}}=&\Delta_n\ln\gamma_2(\h y-\t y),\\
\Delta_m\frac{\t{\t x}\t y(x\t{\t y}+\t x\t{\t{\t y}}+p\t{\t x}\t
y)}{[p(\t xy+\t{\t x}\t y)+x\t{\t y}][p(\t{\t x}\t y+\t{\t{\t
x}}\t{\t y})+\t x\t{\t{\t y}}]}=&\Delta_n\frac{\t xy\t y}{(\h y-\t
y)[p(\t xy+\t{\t x}\t y)+x\t{\t y}]}.
\end{align}
\end{subequations}

\end{appendix}


\end{document}